\documentclass[preprint,floatfix] {revtex4} 
 
\usepackage{graphicx}
\usepackage{subfigure}
\begin{document}

\title{Accurate ro-vibrational spectroscopy of diatomic molecules in a Morse oscillator potential}
\author{Amlan K. Roy}
\altaffiliation{Email: akroy@iiserkol.ac.in, akroy6k@gmail.com}
\affiliation{Division of Chemical Sciences,   
Indian Institute of Science Education and Research (IISER)-Kolkata, 
Mohanpur Campus, P. O. BCKV Campus Main Office, Nadia, 741252, WB, India}

\begin{abstract}
This work presents the bound-state spectra of Morse oscillator, which remains one of the oldest important model potentials for molecules. 
Accurate ro-vibrational energies are obtained by means of a generalized pseudospectral method that offers an optimal, non-uniform 
discretization of the radial grid. Both $s$-wave ($\ell=0$) and rotational ($\ell \neq 0$) states for low and high quantum numbers are calculated 
for four representative diatomic molecules, namely H$_2$, LiH, HCl and CO. First nine states belonging to a maximum of $n, \ell =2$ are 
computed with good accuracy, along with nine other high-lying states for each of these molecules. Present results \emph{surpass} the accuracy of 
\emph{all} hitherto published calculations found so far, except the tridiagonal J-matrix method, which produces similar accuracy as ours. 
Detailed variation of energies with respect to state indices $n,\ell$ show interesting behavior. A host of new states including the higher 
ones are reported as well. This offers a simple general efficient scheme for calculating these and other similar potentials in molecular physics.   
\end{abstract}
\maketitle

\section{Introduction}
Importance of the exponentially varying Morse potential \cite{morse29} in context of vibration-rotation states of diatomic 
molecules has been realized in an enormous amount of work ever since its inception about 85 years ago. The celebrated empirical 
potential has witnessed many important applications in various branches, such as molecular physics, solid-state physics, chemical 
physics, etc \cite{popov01,dong02}. This potential has been studied also in several other contexts. For example, its generalized and 
photon-added coherent states are discussed \cite{dong02a, dong06}; series solution for $D$-dimensional Schr\"odinger equation with 
More potential presented \cite{dong03}; controllability with a compact group SU(2) put forth \cite{dong03a}; a series expansion solution 
of position-dependent mass is given \cite{yu04}; a proper quantization rule where the energy spectra is determined only 
from its ground-state energy has been proposed \cite{serrano10}; also its momentum representation by hypergeometric function is 
suggested \cite{dong12}. It is well known that, exact analytical results are available only for $\ell=0$ states, while the rotational Morse 
oscillator states must be approximated. Thus over the years, a large number of attractive efficient methodologies have been 
developed for better understanding of this model potential employing a multitude of approximations. 

One of the earliest definitive attempts to estimate the accurate solution of this physically significant potential was made in \cite{duff78}. 
Through a modified Morse technique, reasonably good approximate semi-analytic eigenstates were reported for both non-rotating 
and rotating cases in H$_2$, which were complemented by finite-element results. Thereafter, numerous attempts have been made, 
introducing a variety of analytical approximations as well as numerical techniques. Some of the notable ones are mentioned here. The assumption that 
the effective potential for a rotating Morse potential always retains a Morse form, was used in \cite{elsum82} to derive better approximate 
analytic expressions for energy levels, wave functions and other relevant matrix elements in H$_2$. A shifted 1/N expansion 
\cite{morales89} was used for eigenvalues in H$_2$. Later, a modified shifted large-N approach \cite{bag92} restored exact
analytical expressions of $\ell=0$ states and produced quite decent results for $\ell \ne 0$ eigenstates in H$_2$, HCl, CO and LiH molecules.   
A variational method \cite{filho00} in the context of super-symmetric quantum mechanics was proposed to obtain energies and 
eigenfunctions of lowest levels in these molecules. A super-symmetry improvement \cite{morales04} of the Pekeris approximation 
\cite{pekeris34} allowing one to obtain a higher ro-vibrational state from a lower ro-vibrational state, was successfully 
employed for H$_2$. Bound-state eigenvalues and eigenfunctions of diatomic molecular systems in the rotating Morse oscillator
have been presented \cite{berkdemir05} by a combination of Nikiforov-Uvarov method and Pekeris approximation as well. 
In another development \cite{castro06}, analytical solution for eigenstates in H$_2$ molecule was provided by means of a 
two-point quasi-rational approximant technique. Also, the use of a Pekeris approximation within asymptotic iteration method has been 
advocated for eigenvalues \cite{bayrak06,aldossary07} and eigenfunctions \cite{bayrak06} for above diatomic molecules.
Further, through a tridiagonal J-matrix representation \cite{nasser07}, quite impressive results for arbitrary $\ell$ states of Morse 
potential have been reported. A quantization rule \cite{qiang07} within the framework of Pekeris approximation has offered analytical bound 
states for arbitrary $\{n,\ell\}$ quantum numbers in HCl, CO and LiH molecules. In another attempt, approximate analytic bound 
states for generalized q-deformed Morse oscillator were elegantly obtained by a parametric generalization \cite{ikhdair09} of the
Nikiforov-Uvarov method in conjunction with Pekeris approximation.

It is well known that the radial Schr\"odinger equation of this potential can be solved exactly for $\ell=0$ states only. Moreover, several
highly accurate methods are available for pure vibrational levels; e.g., confining the system in a spherical box of certain finite 
radius \cite{leykoo95, taseli98}, asymptotic iteration method \cite{barakat06}, besides the original prescription of Morse 
\cite{morse29}. However for the rotating Morse scenario, exact explicit expressions of eigenvalues and eigenfunctions are not available
in closed form. While some of the above mentioned approximate analytic, semi-analytic or numerical methods produce good quality results for 
$\ell=0$ states, only a very few of these can offer same for arbitrary $\ell \neq 0$ states. Additionally, much attention has been paid 
for ground states, while excited states are reported less frequently and definitively. Moreover, while all these methods consider eigenvalues, 
eigenfunctions have received much less attention relatively. Thus, a simple general prescription which could deliver physically 
meaningful and accurate results for low as well as higher states (eigenvalues and eigenfunctions) in both $\ell=0$ and $\ell \neq 0$ situations, 
would be highly desirable. In this Letter, we examine such a possibility through the generalized pseudospectral (GPS) method. 
This approach has been very successful for a variety of potentials of physical interest, including the spiked harmonic oscillator, rational, 
logarithmic, power-law, Hulth\'en, Yukawa as well as for atoms, molecules with Coulombic singularity, etc., as documented in the following 
references \cite{roy04, roy04b, roy05, roy05a, roy07, roy08, roy08a, roy11, roy13}. The purpose of this Letter is, therefore, 
two-fold: (a) to make a detailed study on the bound-state spectra of Morse potential with special reference to diatomic molecules for 
arbitrary $\{n,\ell\}$ quantum numbers, (b) assess the validity, applicability and performance of GPS method for molecular potentials, 
which has not been done before. To this end, accurate ro-vibrational spectra 
is presented for four diatomic molecules (H$_2$, LiH, HCl, CO) with much emphasis on the energy variations. A thorough comparison 
with literature data has been made wherever possible. The Letter is organized as follows. Section II presents the essential details 
of our methodology employed here. A discussion of our results is given in Section 3 while Section 4 makes a few concluding remarks. 

\section{The GPS method for Morse potential}
In this section, we present an overview of the GPS method employed to solve the radial eigenvalue 
problem with Morse potential. As it has been described before \cite{roy04,roy04b,roy05,roy05a,roy07,roy08,roy08a,roy11, roy13}, here
we provide only the essential steps; details could be found in the references above. 

The desired radial Schr\"odinger equation can be written in the following operator form, 
\begin{equation}
\hat{H}(r)\ \psi(r) = E \ \psi(r),
\end{equation}
where the Hamiltonian operator includes usual kinetic and potential energy operators,  
\begin{equation}
\hat{H}(r) =-\frac{\hbar^2}{2\mu} \frac{d^2}{dr^2} +v(r),  \ \ \ \ \ \mathrm{with} \ \   
v(r) = \frac{\ell (\ell+1)\hbar^2}{2\mu r^2} + D_e \left( e^{-2\alpha x}-2e^{-\alpha x} \right). 
\end{equation}
Here $\ell$ denotes the usual angular momentum quantum number identifying rotational states.  
Last term in Eq.~(2) represents the Morse potential. Potential strength $D_e > 0$, $x=(r-r_e)/r_e$, and $\alpha, r_e$
are two positive parameters. 

A large number of numerical schemes for solution of above radial Schr\"odinger equation employ finite-difference discretization. These often 
require a significant number of spatial grid points chiefly because of their uniform distributional nature. In the GPS method, however, 
a nonuniform optimal spatial discretization could be achieved efficiently as detailed below, maintaining similar kind of good accuracy at 
both small and large $r$ regions. Therefore it requires much lesser grid points to reach convergence, compared to many other methods in the 
literature. One can have a denser mesh at smaller $r$ while a coarser mesh at large $r$ regions. A primary feature of this scheme is that a 
function $f(x)$ in the interval $x \in [-1,1]$ can be approximated by an N-th order polynomial $f_N(x)$ as given below, 
\begin{equation}
f(x) \cong f_N(x) = \sum_{j=0}^{N} f(x_j)\ g_j(x).
\end{equation}
This approximation is \emph {exact} at the \emph {collocation points} $x_j$, i.e., $f_N(x_j) = f(x_j)$. In the Legendre pseudospectral 
method used in this work we also have, $x_0=-1$, $x_N=1$, while $x_j (j=1,\ldots,N-1)$ are obtained from roots of first derivative of 
$P_N(x)$ with respect to $x$, i.e., $P'_N(x_j) = 0$. The cardinal functions $g_j(x)$ in Eq.~(3) are given by following expression,
\begin{equation}
g_j(x) = -\frac{1}{N(N+1)P_N(x_j)}\ \  \frac{(1-x^2)\ P'_N(x)}{x-x_j},
\end{equation}
thereby satisfying a unique property $g_j(x_{j'}) = \delta_{j'j}$. The semi-infinite domain $r \in [0, \infty]$ is now mapped into 
a finite domain $x \in [-1,1]$ by employing the transformation $r=r(x)$. Now invoking an algebraic nonlinear mapping of the form, 
$ r=r(x)=L\ \ \frac{1+x}{1-x+\alpha},$ 
with L and $\alpha=2L/r_{max}$ as mapping parameters, and introducing a symmetrization procedure leads to a transformed 
Hamiltonian as below, 
\begin{equation}
\hat{H}(x)= -\frac{1}{2} \ \frac{1}{r'(x)}\ \frac{d^2}{dx^2} \ \frac{1}{r'(x)}
+ v(r(x))+v_m(x), \ \ \ \ \ \ \ \ v_m(x)=\frac {3(r'')^2-2r'''r'}{8(r')^4}.
\end{equation}
The advantage is that this leads to a  \emph {symmetric} matrix eigenvalue problem which can be readily solved to obtain 
accurate eigenvalues and eigenfunctions. For the particular transformation used above, $v_m(x)=0$. So, finally one obtains the following
set of coupled equations, 
\begin{widetext}
\begin{equation}
\sum_{j=0}^N \left[ -\frac{1}{2} D^{(2)}_{j'j} + \delta_{j'j} \ v(r(x_j))
+\delta_{j'j}\ v_m(r(x_j))\right] A_j = EA_{j'},\ \ \ \ j=1,\ldots,N-1,
\end{equation}
\end{widetext}
where
\begin{equation}
A_j  = \left[ r'(x_j)\right]^{1/2} \psi(r(x_j))\ \left[ P_N(x_j)\right]^{-1}, \ \ \ \ \ 
D^{(2)}_{j'j} =  \left[r'(x_{j'}) \right]^{-1} d^{(2)}_{j'j} 
\left[r'(x_j)\right]^{-1}, 
\end{equation}
with
\begin{eqnarray}
d^{(2)}_{j',j} & = & \frac{1}{r'(x)} \ \frac{(N+1)(N+2)} {6(1-x_j)^2} \ 
\frac{1}{r'(x)}, \ \ \ j=j', \nonumber \\
 & & \nonumber \\
& = & \frac{1}{r'(x_{j'})} \ \ \frac{1}{(x_j-x_{j'})^2} \ \frac{1}{r'(x_j)}, 
\ \ \ j\neq j'.
\end{eqnarray}
A sufficiently large number of test calculations were done to check the performance of this method with respect to the mapping parameters. 
Results are 
reported here only up to the precision that were found to maintain stability with such variations. In this way, 
a consistent set of numerical parameters of $\alpha=25$, $N=300$ were chosen, which seemed to be appropriate and satisfactory for 
our purposes. In general, $r_{max}=200$ a.u. was sufficient for most calculations, but for higher excited states, it needed to be 
increased to capture the long-range tails in wave function.

\begingroup
\squeezetable
\begin{table}
\caption {\label{tab:table1}Spectroscopic parameters of selected molecules, used in the present calculation.}
\begin{ruledtabular}
\begin{tabular}{lllll}
Molecule  &  $D_e$ (eV)  & $r_e$ ($\AA$) & $\mu$ (amu) &  $\alpha$  \\
\hline
   H$_2$ \cite{nasser07}    & 4.7446    & 0.7416  & 0.50391    & 1.440558        \\ 
 LiH \cite{nasser07}        & 2.515287  & 1.5956  & 0.8801221  & 1.7998368       \\ 
 HCl \cite{nasser07}        & 4.61907   & 1.2746  & 0.9801045  & 2.38057         \\
 CO \cite{nasser07}         & 11.2256   & 1.1283  & 6.8606719  & 2.59441         \\  
\end{tabular}
\end{ruledtabular}
\end{table}
\endgroup

\section{Results and Discussion}
The computed ro-vibrational energy levels of diatomic molecules are now examined. For convenience of comparison, we choose four molecules, 
\emph{viz.}, H$_2$, LiH, HCl, CO, which have been most widely studied in the literature. The model parameters, as quoted in \cite{nasser07}, 
are adopted for all the results reported in this work. The following conversion factors from NIST database \cite{nist} are used: Bohr 
radius = 0.52917721092 \AA, Hartree energy = 27.21138505 eV, and electron rest mass = 5.48577990946 $\times 10^{-4}$ u. Table 2 compares 
all the nine calculated bound states of Morse potential corresponding to $\{n,\ell\}$ quantum numbers 0, 1 and 2, with available 
literature data. For H$_2$, a host of results are available for all the $\ell$ states with vibrational quantum number $n=0$, whereas for other 
molecules, same is true for only the non-rotational case ($\ell=0$) having $n=0$. For all these molecules, however, no results could be found 
for states with both non-zero $n, \ell$ quantum numbers for direct comparison. Note that, $s$-wave eigenstates offer exact analytical results 
and thus can be used as a valuable guide to assess the quality of our computed energies. It is observed that, GPS results for all $\ell=0$ 
states match perfectly with the exact results. As these are available in \cite{nasser07}, they are not quoted here again to avoid repetition.  
The semi-analytic modified Morse results \cite{duff78} for $n=0$ in H$_2$, worsen in quality with an increase in $\ell$. Same energy for the 
lowest state of H$_2$ was also found in the approximate analytic works of \cite{depristo81,elsum82}. The lowest state of H$_2$ has also been 
studied by means of shifted 1/N expansion method \cite{morales89}, hierarchical super-symmetric improvement \cite{morales04} of a Pekeris 
approximation \cite{pekeris34}, two-point quasi-rational approximation \cite{castro06}, and other some methods with moderate success. 
Energies from the modified shifted large-N approach \cite{bag92}, asymptotic iteration method \cite{aldossary07,bayrak06} in conjunction with 
the original prescription of Pekeris, a variational method \cite{filho00} within super-symmetric quantum mechanics, a combination of the parametric 
generalization of Nikiforov-Uvarov method \cite{ikhdair09} with Pekeris scheme, etc., have also been reasonably good for $n=\ell=0$ state of these 
molecules. The lowest states of LiH, HCl and CO were also obtained nicely by means of an exact quantization rule \cite{qiang07} in the 
Pekeris framework. However, it seems, wherever available, amongst all the methods quoted here, the tridiagonal J-matrix \cite{nasser07} provides 
the best reference energies for rotational Morse oscillator. For H$_2$, LiH, they employed (100,40), (100,60) Laguerre bases while for HCl, CO, 
they found (100,5), (200,12) oscillator bases to be more suitable. It is quite gratifying that, for all the states, our present GPS results exactly 
reproduce these accurate eigenvalues. 

\begingroup
\squeezetable
\begin{table}
\caption {\label{tab:table2}Calculated negative eigenvalues (in eV) of Morse potential for
$\{n,\ell\}=0,1,2$ states of some diatomic molecules along with literature data. PR signifies Present Result.} 
\begin{ruledtabular}
\begin{tabular}{llll|llll}
$n$  &  $\ell$  & $-$E (PR)   & $-$E (Literature) &  $n$  &  $\ell$  & $-$E (PR)   & $-$E (Literature) \\
\hline
    &    &   \underline{H$_2$}   &    &   &          &    \underline{LiH}     &  \\
0   & 0  & 4.47601313  & 4.4760\footnotemark[1]$^,$\footnotemark[2]$^,$\footnotemark[3]$^,$\footnotemark[4],
4.4749\footnotemark[5]$^,$\footnotemark[10],4.4758\footnotemark[6], & 
0   & 0  & 2.42886321  & 2.4280\footnotemark[4],2.4291\footnotemark[6],2.4389\footnotemark[8],2.4278\footnotemark[10], \\
    &    &             & 4.47601\footnotemark[7]$^,$\footnotemark[8]$^,$\footnotemark[12],4.4760084\footnotemark[9],4.47601313\footnotemark[11] 
    &    &   &   & 2.42886321\footnotemark[11],2.42886\footnotemark[12]$^,$\footnotemark[14],2.4287\footnotemark[13] \\
    & 1  & 4.46122852  & 4.4618\footnotemark[1],4.4612233\footnotemark[9]  &     & 1  & 2.42702210   &    \\ 
    & 2  & 4.43179975  & 4.4335\footnotemark[1],4.4318\footnotemark[2],4.4317934\footnotemark[9]  &     & 2  & 2.42334244   &    \\ 
1   & 0  & 3.96231534  & 3.96231535\footnotemark[11]  &  1  & 0  & 2.26054805   &  2.26054805\footnotemark[11]   \\
    & 1  & 3.94811647  &   &     & 1  & 2.25875559   &    \\
    & 2  & 3.91986423  &   &     & 2  & 2.25517324   &    \\
2   & 0  & 3.47991882  & 3.47991884\footnotemark[11]  &  2  & 0  & 2.09827611   &  2.09827611\footnotemark[11]  \\
    & 1  & 3.46633875  &   &     & 1  & 2.09653304   &    \\
    & 2  & 3.43932836  &   &     & 2  & 2.09304950   &    \\
\hline
    &    &   \underline{HCl}     &    &   &          &    \underline{CO}     &  \\
0   & 0  & 4.43556394  & 4.4355\footnotemark[4],4.4360\footnotemark[6],4.4356\footnotemark[8],    &  
0   & 0  & 11.09153532 & 11.092\footnotemark[4],11.093\footnotemark[6],11.0915\footnotemark[8]$^,$\footnotemark[12]$^,$\footnotemark[14], \\
    &    &             & 4.4352\footnotemark[10],4.43556394\footnotemark[11],4.43556\footnotemark[12]$^,$\footnotemark[14]  &   &   &    
    & 11.091\footnotemark[10]$^,$\footnotemark[13],11.09153532\footnotemark[11]  \\
    & 1  & 4.43297753  &   &     & 1  & 11.09105875  &    \\ 
    & 2  & 4.42780630  &   &     & 2  & 11.09010565  &    \\ 
1   & 0  & 4.07971006  & 4.07971007\footnotemark[11]  &  1  & 0  & 10.82582206  & 10.82582207\footnotemark[11]   \\ 
    & 1  & 4.07720144  &   &     & 1  & 10.82534959  &    \\ 
    & 2  & 4.07218579  &   &     & 2  & 10.82440465  &    \\ 
2   & 0  & 3.73873384  & 3.73873385\footnotemark[11]  &  2  & 0  & 10.56333028  & 10.56333029\footnotemark[11]   \\ 
    & 1  & 3.73630382  &   &     & 1  & 10.56286190  &    \\ 
    & 2  & 3.73144539  &   &     & 2  & 10.56192516  &    \\ 
\end{tabular}
\end{ruledtabular}
\begin{tabbing}
$^{\mathrm{a}}$Ref.~\cite{duff78}. \hspace{27pt}  \=
$^{\mathrm{b}}$Ref.~\cite{depristo81}. \hspace{27pt}  \=
$^{\mathrm{c}}$Ref.~\cite{elsum82}. \hspace{27pt}  \=
$^{\mathrm{d}}$Ref.~\cite{bag92}. \hspace{27pt}  \=
$^{\mathrm{e}}$Ref.~\cite{morales89}. \hspace{27pt}  \=
$^{\mathrm{f}}$Ref.~\cite{filho00}. \hspace{27pt}  \=
$^{\mathrm{g}}$Ref.~\cite{morales04}.                \\
$^{\mathrm{h}}$Ref.~\cite{bayrak06}. \hspace{27pt}  \=
$^{\mathrm{i}}$Ref.~\cite{castro06}. \hspace{27pt}  \=
$^{\mathrm{j}}$Ref.~\cite{aldossary07}. \hspace{27pt}  \=
$^{\mathrm{k}}$Ref.~\cite{nasser07}. \hspace{27pt}  \=
$^{\mathrm{l}}$Ref.~\cite{ikhdair09}. \hspace{27pt}  \=
$^{\mathrm{m}}$Ref.~\cite{berkdemir05}. \hspace{27pt}  \=
$^{\mathrm{n}}$Ref.~\cite{qiang07}. 
\end{tabbing}
\end{table}
\endgroup

As a further test of the performance of GPS scheme, Table 3 compares energies of some high-lying states ($\ell=10, 20, 25; n=0-5$) 
of H$_2$, LiH, HCl and CO with literature results. A decent number of reference energies exist in this case (more studies were made 
on H$_2$ than the other three) for $\ell=10$, and some of them are quoted. However, for $\ell=20$, only few reliable results are available
and none can be found for $\ell=25$. The semi-analytic eigenvalues \cite{duff78,depristo81} were among some of the very first estimates 
for $\ell=10$ states with $n=0$ and 5 in H$_2$; although much improved results exist now. The $n=0,5$ states
with $\ell=10,20$ of H$_2$ were also treated in \cite{elsum82}, which tacitly assumes that the effective potential for a rotating Morse 
oscillator always retains a Morse form; a shifted 1/N expansion \cite{morales89}, a combination of Pekeris formalism and super-symmetric 
quantum mechanics \cite{morales04} producing very similar quality eigenvalues. The $(0,10)$ and $(0,20)$ state energies of H$_2$ were also 
calculated by a two-point quasi-rational approximation \cite{castro06}. Here two numbers in parentheses denote $n$, $\ell$ quantum numbers 
respectively. For all these molecules, first systematic energies and eigenfunctions of $\ell=10$ states having $n=0,5$ were reported by a 
modified shifted large-N approach \cite{bag92}, producing nice quality results. By applying a variational method in super-symmetric quantum 
mechanics \cite{filho00}, moderate estimates of (0,10) states of all these molecules were presented. The (0,10) and (5,10) 
eigenstates of these molecules were also estimated with reasonable success by asymptotic iteration method and some of its variants 
\cite{bayrak06, aldossary07}, as well by a parametric generalization of Nikiforov-Uvarov method \cite{ikhdair09}. The (0,10) and 
(5,10) states of LiH and CO are available from the combined Nikiforov-Uvarov method and Pekeris approach \cite{berkdemir05}. The $n=0,5$ 
vibrational states corresponding to rotational quantum number $\ell=10, 20$ have been reported via an exact quantization route \cite{qiang07}
coupled with the Pekeris approximation, for LiH, HCl and CO. However, as in Table 1, in this case also, it appears that the most accurate energies 
are offered by J-matrix basis that facilitates a tridiagonal representation of the reference Hamiltonian \cite{nasser07}. For H$_2$ and CO, 
it is seen that agreement of present eigenvalues with that of \cite{nasser07} is excellent (two results are virtually identical; absolute
deviation remains only $10^{-8}$ eV consistently for all eight states under discussion, except for E$_{5,20}$ of CO, where an exact 
matching is observed). Unexpectedly however, for LiH and HCl, these two energies show significant deviation from each other. Part of the reason 
may be rooted in the fact that, the $s$-wave ($\ell=0$) and rotational ($\ell \ne 0$) bound states of Morse potential were estimated by 
same parameter sets for H$_2$ ($N=100, \lambda =40$ in Tables 2 and 3 in \cite{nasser07}) and CO ($N=200, \lambda=12$ in Tables 2 and 6 in 
\cite{nasser07}). However, $\ell=0$ and $\ell \neq 0$ states of HCl were calculated by two different parameter sets in Table 2 ($N=100, 
\lambda=5$) and Table 5 ($N=100, \lambda =6$) in \cite{nasser07}. From a careful observation of these two tables in \cite{nasser07}, it is 
clear that the three common $s$-wave energies (namely E$_{0,0}$, E$_{5,0}$, E$_{7,0}$) differ from each other with respect to changes in the 
parameter $\lambda$. Since our $s$-wave results practically coincide with the exact results (quoted in Table 2 of \cite{nasser07}) and also with 
their own calculations in Table 2, the slight differences in our results from those in \cite{nasser07} for the rotational case of HCl may be 
attributed to the different parameter sets employed by them. We anticipated a similar phenomenon for LiH. However, the three common $l=0$ 
state-energies \emph{viz.}, E$_{0,0}$, E$_{5,0}$, E$_{7,0}$ of LiH in reference \cite{nasser07}, differ from each other in Tables 2 and 4, 
even though same parameter set ($N=100, \lambda=60$) was employed in both tables. It is not clear to us if there is any other factor responsible 
for this discrepancy. In any case, keeping in mind the excellent accuracy and reliability GPS method has offered for all the above states in 
various molecules and also in many other previous situations \cite{roy04,roy04b,roy05,roy05a,roy07,roy08,roy08a,roy11,roy13}, we are led to 
conclude that the results for LiH are also as accurate as for the other three molecules. 

\begingroup
\squeezetable
\begin{table}
\caption {\label{tab:table3}Calculated negative eigenvalues (in eV) of some selected high-lying states of Morse potential with $n=0-5; \ell=10,20,25$ 
for some diatomic molecules along with literature data. PR signifies Present Result.} 
\begin{ruledtabular}
\begin{tabular}{llll|llll}
$n$  &  $\ell$  & $-$E (PR)   & $-$E (Literature) &  $n$  &  $\ell$  & $-$E (PR)   & $-$E (Literature) \\
\hline
    &    &   \underline{H$_2$}   &    &   &          &    \underline{LiH}     &  \\
0   & 10 & 3.7247470  & 3.7506\footnotemark[1],3.7250\footnotemark[2],3.7249\footnotemark[3],
3.7247\footnotemark[4]$^,$\footnotemark[5]$^,$\footnotemark[10],3.7187\footnotemark[6],      &  
0   & 10 & 2.3288546  & 2.3261\footnotemark[5]$^,$\footnotemark[10],2.3287\footnotemark[6]$^,$\footnotemark[13],2.3288\footnotemark[8],   \\ 
    &    &            & 3.72193\footnotemark[7]$^,$\footnotemark[8],3.7247181\footnotemark[9],3.7247471\footnotemark[11],3.72194\footnotemark[12]  
&   &  &  & 2.3288530\footnotemark[11],2.32884\footnotemark[12],2.32883\footnotemark[14]   \\
3   &    & 2.3833482  &   &  3  &     & 1.8502014    &    \\ 
5   &    & 1.6526901  & 1.849\footnotemark[1],1.657\footnotemark[2],1.6523\footnotemark[3],1.6526\footnotemark[4]$^,$\footnotemark[10],
1.6535\footnotemark[5],   &  
5   &    & 1.5615170  & 1.5525\footnotemark[5],1.5607\footnotemark[8],1.5479\footnotemark[10],    \\
    &    &            & 1.64002\footnotemark[7],1.60391\footnotemark[8]$^,$\footnotemark[12],1.6526902\footnotemark[11]   &  &  &  
    & 1.5615114\footnotemark[11],1.56074\footnotemark[12]$^,$\footnotemark[14],1.5606\footnotemark[13] \\
0   & 20 & 2.0840635  & 2.0841\footnotemark[3],2.0839\footnotemark[4],2.0735\footnotemark[6],2.02864\footnotemark[7], &  0  & 20   
    & 2.0600120       & 2.0600073\footnotemark[11],2.05977\footnotemark[14]   \\ 
    &    &            & 2.0839937\footnotemark[9],2.0840636\footnotemark[11]        &    &   &     &      \\
3   &    & 1.0423209  &   &  3  &     & 1.6044463    &    \\ 
5   &    & 0.5237656  & 0.5093\footnotemark[3],0.5233\footnotemark[4],0.42101\footnotemark[7],0.5237657\footnotemark[11]   &  5  &     
         & 1.3316820  & 1.3316742\footnotemark[11],1.32718\footnotemark[14]    \\ 
0   & 25 & 1.1659941  &   &  0  & 25  & 1.8719967    &    \\ 
3   &    & 0.3405278  &   &  3  &     & 1.4335914    &    \\ 
4   &    & 0.1405719  &   &  5  &     & 1.1726358    &    \\ 
\hline
    &    &   \underline{HCl}     &    &   &          &    \underline{CO}     &  \\
0   & 10 & 4.2940924  & 4.2940\footnotemark[5]$^,$\footnotemark[6],4.2941\footnotemark[8],4.2939\footnotemark[10],4.2940628\footnotemark[11],  &  
0   & 10 & 11.0653333 & 11.065\footnotemark[5]$^,$\footnotemark[10]$^,$\footnotemark[13],11.066\footnotemark[6], \\
    &    &            & 4.29408\footnotemark[12],4.29407\footnotemark[14]       &     &     &   
    & 11.0653\footnotemark[8]$^,$\footnotemark[12]$^,$\footnotemark[14],11.0653334\footnotemark[11]   \\ 
3   &    & 3.2841469  &   &  3  &     & 10.2785342   &    \\ 
5   &    & 2.6854833  & 2.6850\footnotemark[5],2.6847\footnotemark[8],2.6712\footnotemark[10],2.6853673\footnotemark[11],  &  
5   &    &  9.7701123 & 9.770\footnotemark[5],9.7701\footnotemark[8],9.765\footnotemark[10],    \\
    &    &            & 2.68471\footnotemark[12],2.68472\footnotemark[14]   &    &    &    
    & 9.7701124\footnotemark[11],9.77009\footnotemark[12],9.769\footnotemark[13],9.77011\footnotemark[14]   \\
0   & 20 & 3.9038526  & 3.9037744\footnotemark[11],3.90374\footnotemark[14]  & 0 & 20 & 10.9915901 
    & 10.9915902\footnotemark[11],10.9916\footnotemark[14]   \\ 
3   &    & 2.9306329  &   &  3  &     & 10.2066975   &    \\ 
5   &    & 2.3571828  & 2.3570303\footnotemark[11],2.35354\footnotemark[14]  &  5  &  &  9.6995563   
    & 9.6995563\footnotemark[11],9.69952\footnotemark[14]   \\ 
0   & 25 & 3.6222352  &   &  0  & 25  & 10.9369716   &    \\ 
3   &    & 2.6764118  &   &  3  &     & 10.1534940   &    \\ 
5   &    & 2.1218117  &   &  5  &     &  9.6473034   &    \\ 
\end{tabular}
\end{ruledtabular}
\begin{tabbing}
$^{\mathrm{a}}$Ref.~\cite{duff78}. \hspace{27pt}  \=
$^{\mathrm{b}}$Ref.~\cite{depristo81}. \hspace{27pt}  \=
$^{\mathrm{c}}$Ref.~\cite{elsum82}. \hspace{27pt}  \=
$^{\mathrm{d}}$Ref.~\cite{morales89}. \hspace{27pt}  \=
$^{\mathrm{e}}$Ref.~\cite{bag92}. \hspace{27pt}  \=
$^{\mathrm{f}}$Ref.~\cite{filho00}. \hspace{27pt}  \=
$^{\mathrm{g}}$Ref.~\cite{morales04}.                \\
$^{\mathrm{h}}$Ref.~\cite{bayrak06}. \hspace{27pt}  \=
$^{\mathrm{i}}$Ref.~\cite{castro06}. \hspace{27pt}  \=
$^{\mathrm{j}}$Ref.~\cite{aldossary07}. \hspace{27pt}  \=
$^{\mathrm{k}}$Ref.~\cite{nasser07}. \hspace{27pt}  \=
$^{\mathrm{l}}$Ref.~\cite{ikhdair09}. \hspace{27pt}  \=
$^{\mathrm{m}}$Ref.~\cite{berkdemir05}. \hspace{27pt}  \=
$^{\mathrm{n}}$Ref.~\cite{qiang07}.
\end{tabbing}
\end{table}
\endgroup

Once the accuracy and reliability of our method is established, we now investigate the variation of energies in different states. For that, in 
Fig.~1, representative plots are given, where (a), (b), (c) in left panel correspond to the changes in bound-state energies (in eV) of H$_2$, LiH, CO, 
with respect to the vibrational quantum number $n$ at six selected $\ell$ values, \emph{viz.}, 0, 5, 10, 15, 20, 25. Figures (d), (e), (f) in 
the right-hand panel signify corresponding energy changes (in eV) of H$_2$, LiH and CO molecules with respect to rotational quantum number
$\ell$ at six selected $n$ values, namely 0, 3, 6, 9, 12 and 15 (for H$_2$, only the first five $n$) respectively. While the $\ell$ axis in 
(d)--(f) has been kept fixed at 25, the $n$ axis in 
(a)--(c), on the other hand, has been varied to capture the necessary qualitative structures in these plots. This is a consequence of the fact that
this potential supports a limited number of bound states for these molecules; the estimated $n_{max}$ for the three molecules under consideration
being 17, 29 and 83 \cite{nasser07, ikhdair09} respectively. Plots for HCl is not presented here as its qualitative features remain quite similar to
that in LiH. To our knowledge, such detailed energy plots have not been presented before except that for LiH in \cite{nasser07} and we hope these 
results would be helpful for future studies. In going from H$_2$ to LiH to CO, it is seen that, E$_{n,\ell}$ versus $n$ plots for different $\ell$
become more closely spaced, with H$_2$ showing maximum sparsity. Also the rate of increase in energy, in general, increases as one moves towards 
H$_2$-LiH-CO, showing almost a linear behavior for CO. On the other hand, E$_{n,\ell}$ versus $\ell$ plots for all three molecules remain 
well separated. As one goes through H$_2$-LiH-CO, however, the plots for a given $n$ seem to vary very slowly, with H$_2$, again showing the 
most rapid increase in energy. For a given molecule, E versus $\ell$ seems to change far less appreciably as $n$ progresses to higher values. 
In both the E$_{n,\ell}$ versus $n$ and $\ell$ cases, however, the individual plots for a given molecule remain nearly parallel to each other. 

\begin{figure}
\centering
\begin{minipage}[t]{0.40\textwidth}\centering
\includegraphics[scale=0.38]{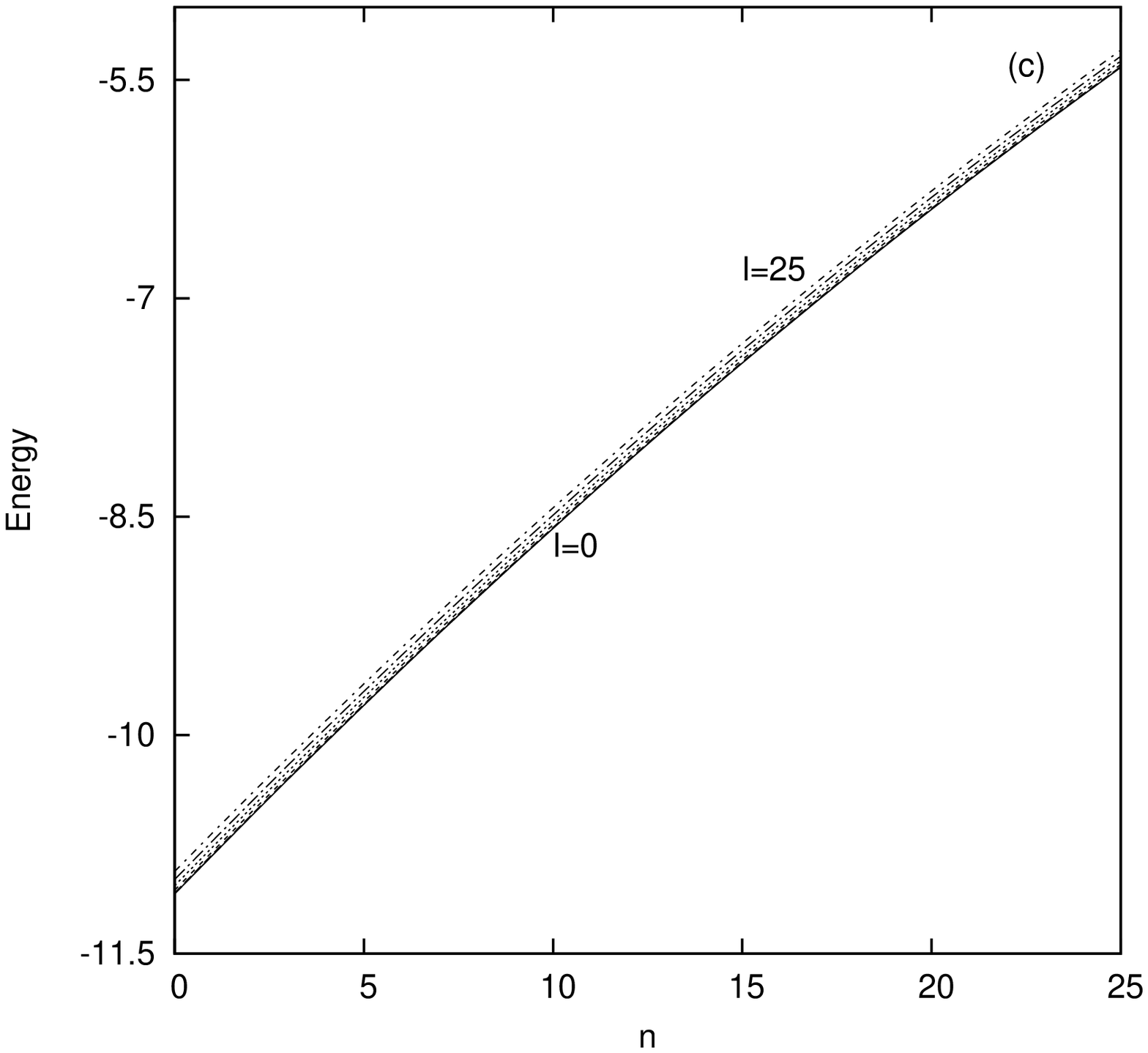}
\end{minipage}
\hspace{0.15in}
\begin{minipage}[t]{0.35\textwidth}\centering
\includegraphics[scale=0.38]{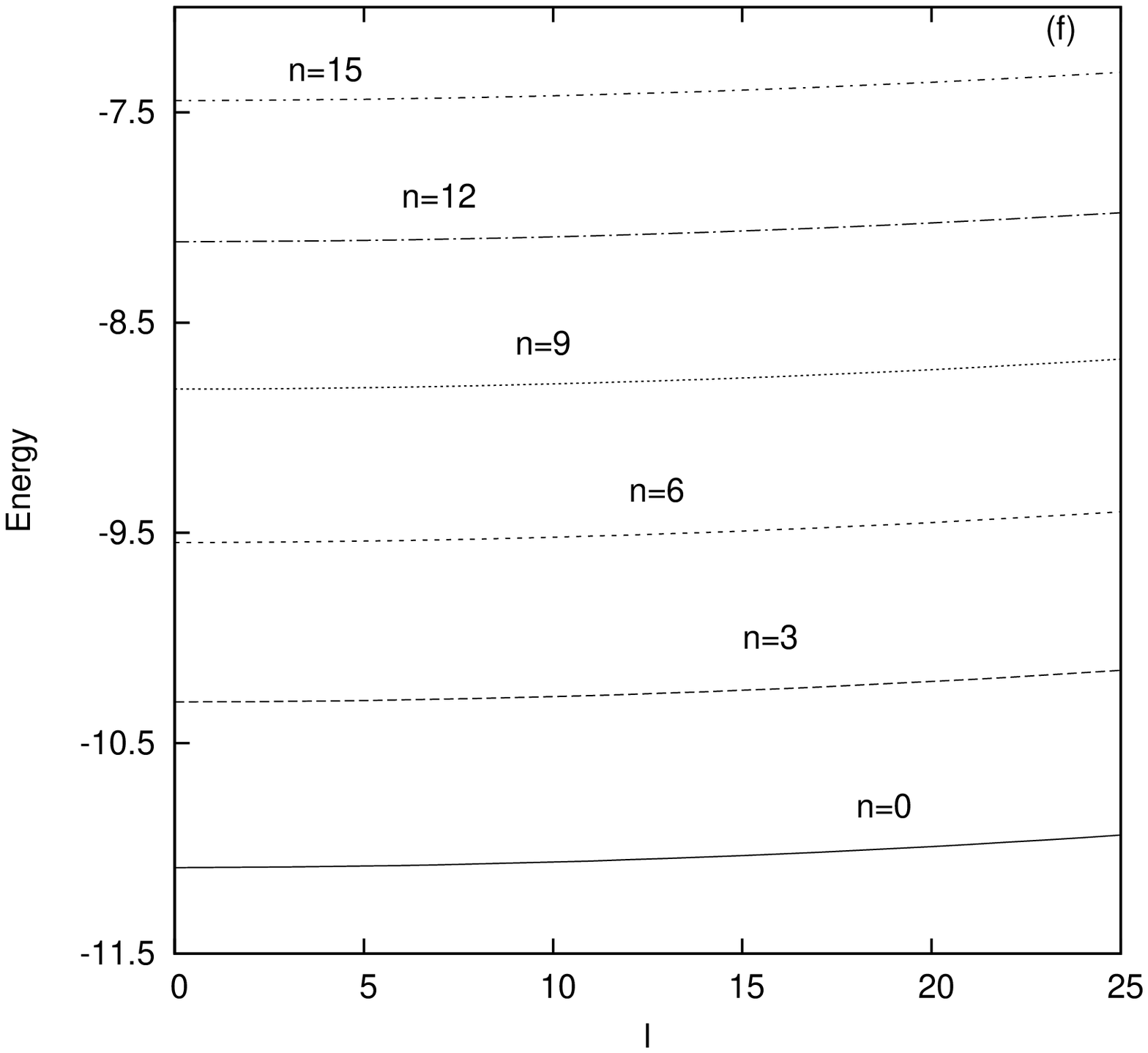}
\end{minipage}
\\[10pt]
\begin{minipage}[t]{0.40\textwidth}\centering
\includegraphics[scale=0.38]{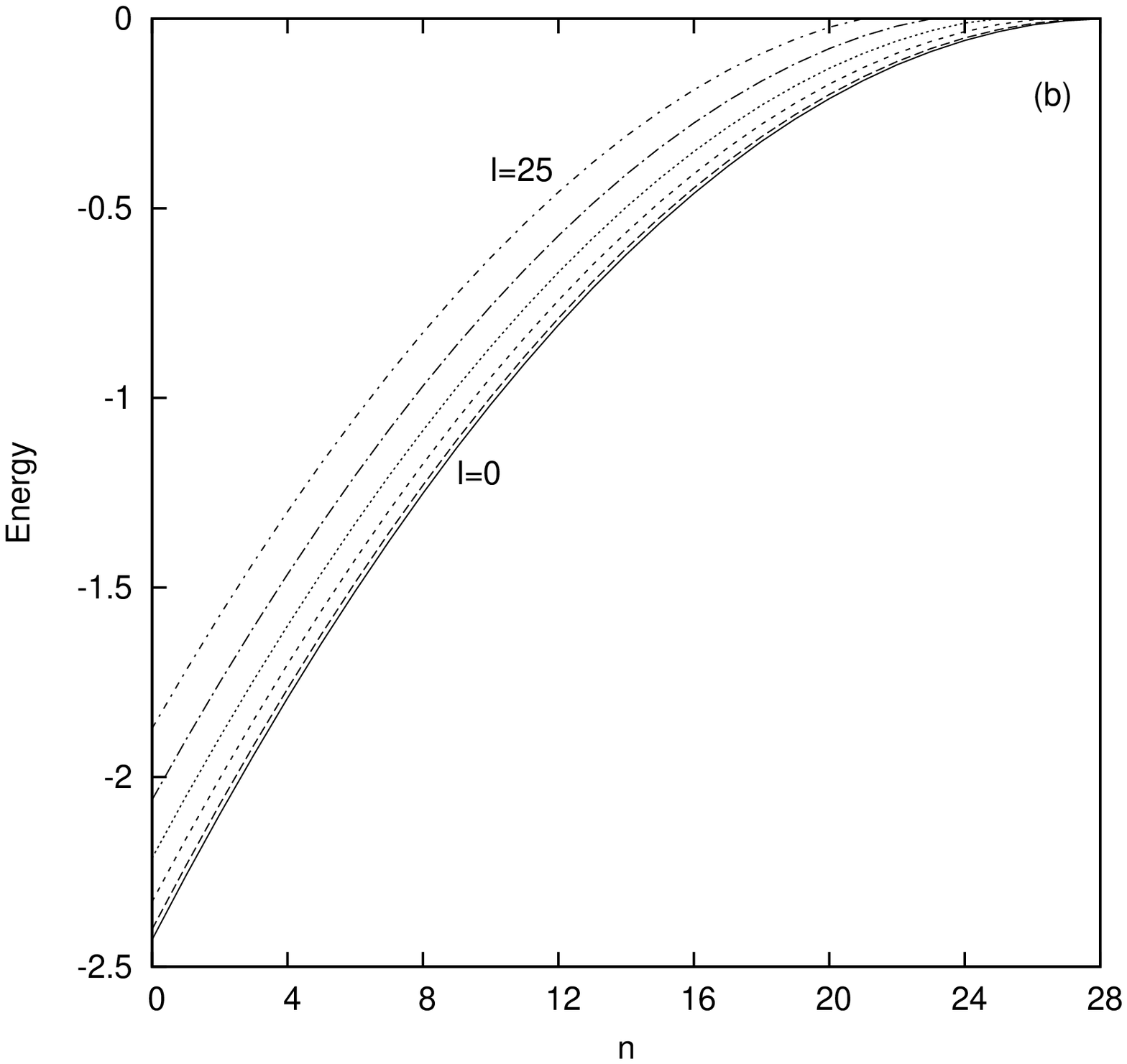}
\end{minipage}
\hspace{0.15in}
\begin{minipage}[t]{0.35\textwidth}\centering
\includegraphics[scale=0.38]{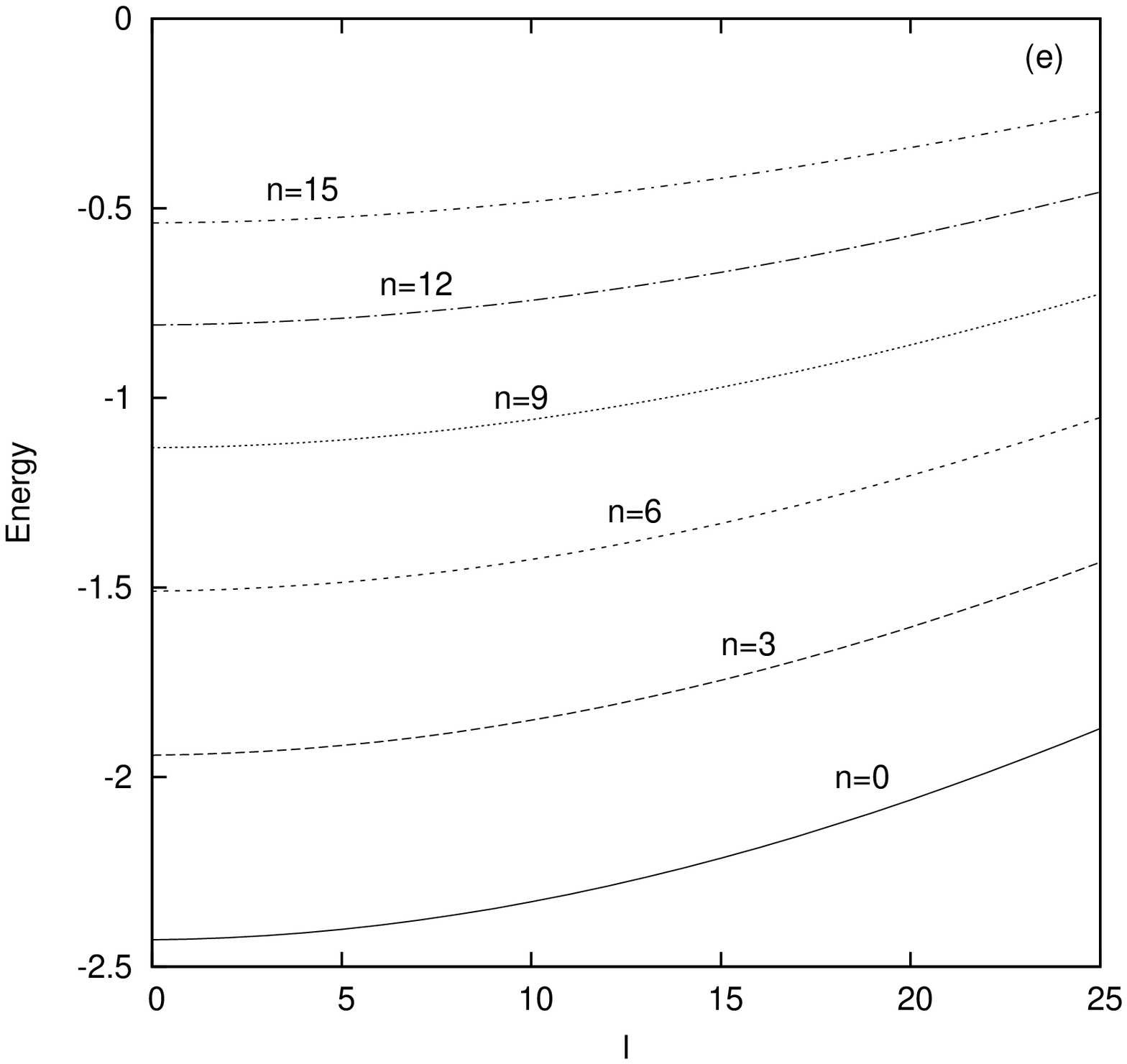}
\end{minipage}
\\[10pt]
\begin{minipage}[b]{0.40\textwidth}\centering
\includegraphics[scale=0.38]{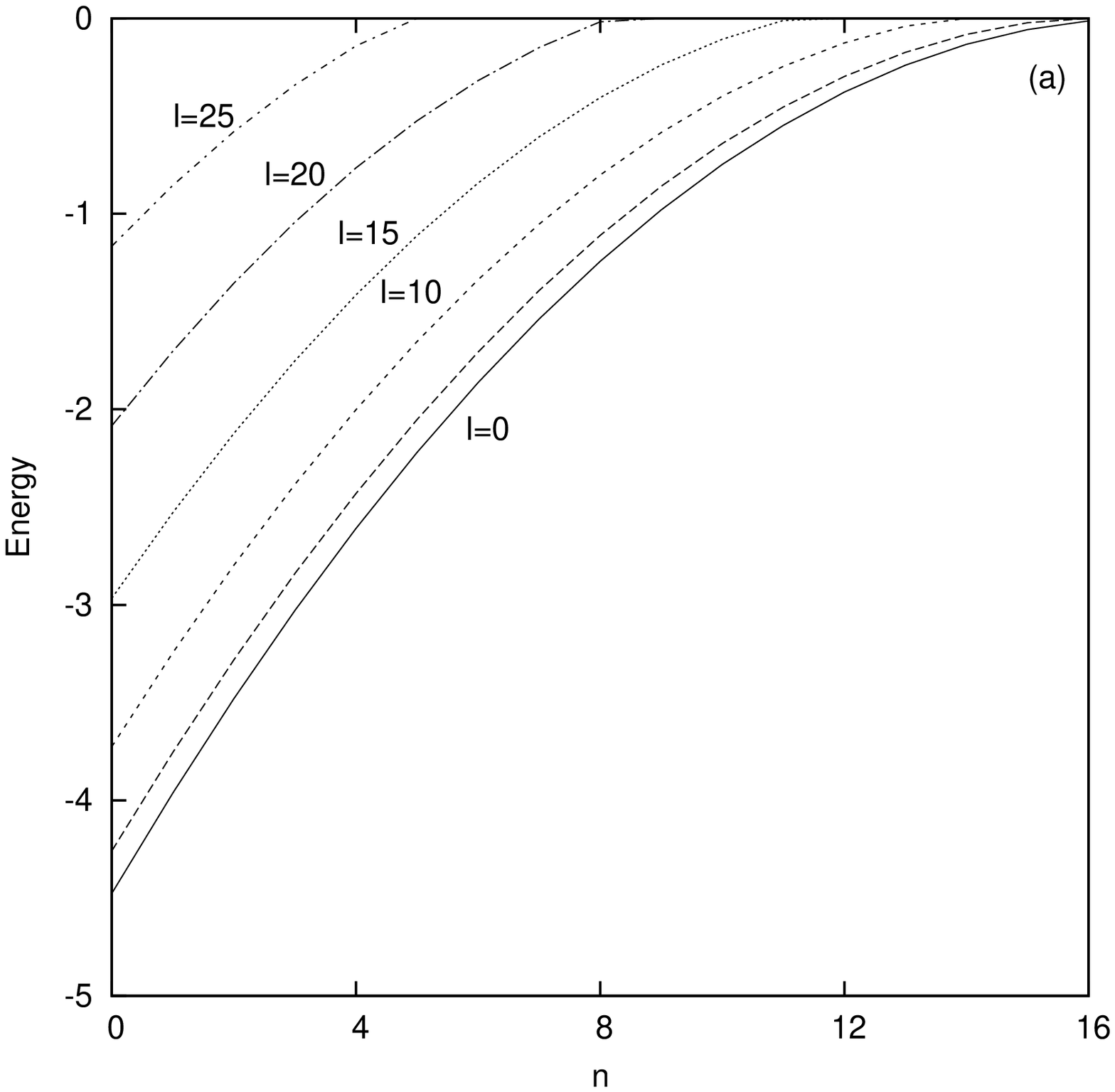}
\end{minipage}
\hspace{0.15in}
\begin{minipage}[b]{0.35\textwidth}\centering
\includegraphics[scale=0.38]{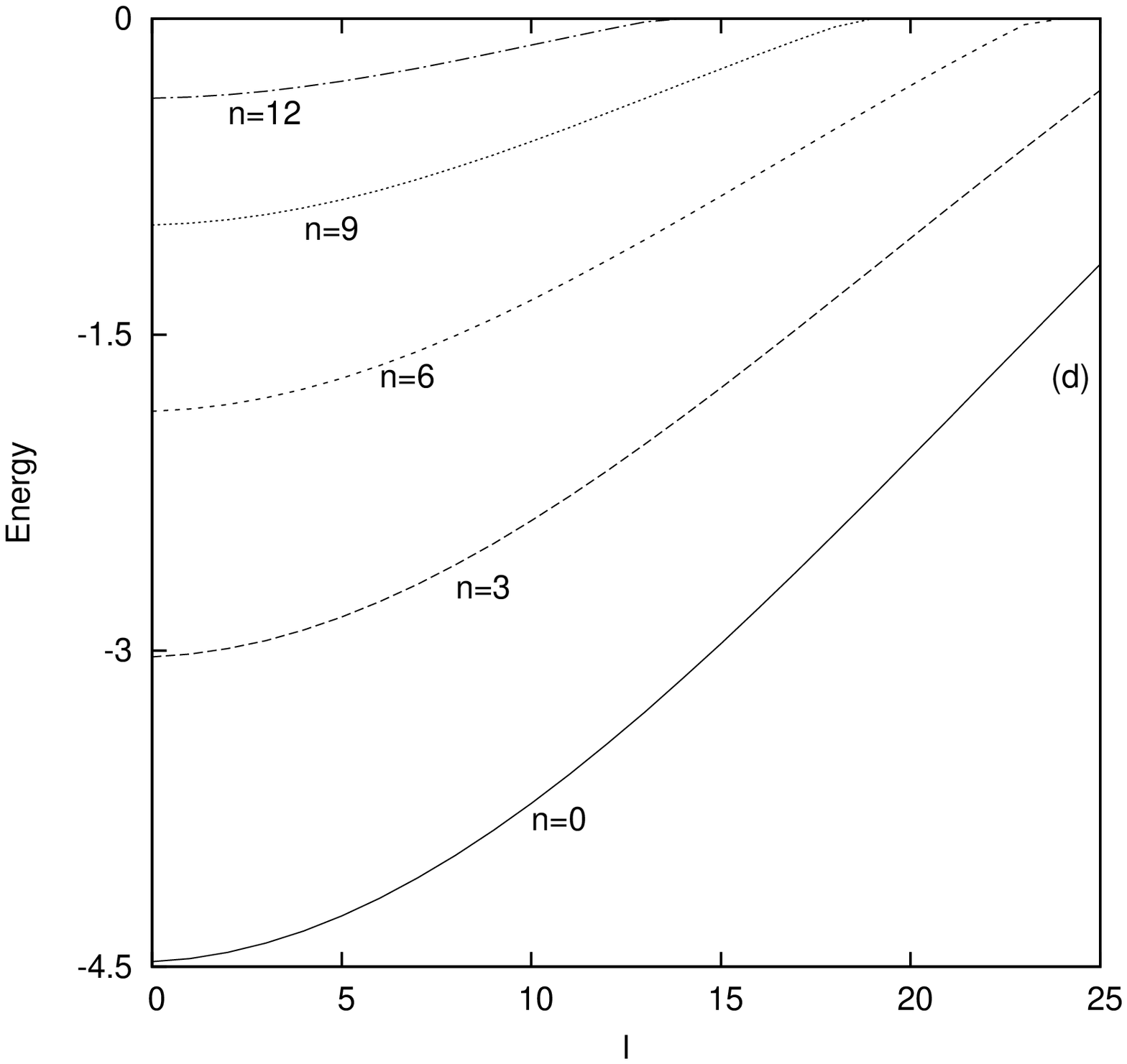}
\end{minipage}
\caption[optional]{Energy variations (in eV) in Morse potential, with respect to vibrational ($n$) (left panel) and rotational ($\ell$) (right panel)
quantum numbers respectively. In the former case, six $\ell$ values of 0, 5 ,10, 15, 20, 25, and for the latter, six $n$ values of 0, 3, 6, 9, 12, 
15 (for H$_2$, only the first five) were selected. (a), (d) correspond to H$_2$; (b), (e) correspond to LiH; and (c), (f) to CO respectively. 
See text for details.}
\end{figure}

At this stage, a few passing comments are in order. As already mentioned, numerous approximate analytical and numerical schemes were
put forth for accurate estimate of eigenstates and eigenvalues of Morse potential over the years. However, some of these are for the non-rotational 
case only or good mainly for lower states; while in yet others, only eigenvalues are obtained. In this work, we have reported low and high vibrational
states of Morse potentials in both $s$-wave and rotational cases with excellent accuracy. The GPS formalism surpasses the results of all
existing methods except the tridiagonal matrix representation approach of \cite{nasser07}. And in most cases, there is near-exact agreement between 
these two. The method is accurate, simple, faithful, and this is true for both non-rotational and rotational cases. Lower and higher states are
obtained in a straightforward manner without any extra difficulty, which is not always possible, in some of the methods in literature. Sometimes, 
slight variations in our energies from references may occur because of the minor differences in conversion factors used by these authors. However, 
that has no bearing on the main objectives of this work. As produced in many of the works 
\cite{roy04,roy04b,roy05,roy05a,roy07,roy08,roy08a,roy11,roy13} before, eigenfunctions are obtained with equal ease in this method, and thus are 
omitted here for brevity. 
 
\section{conclusion}
A detailed investigation has been made on eigenvalues, eigenfunctions of ro-vibrational levels of Morse oscillator by using the GPS method. It  
is simple, easy to use, computationally efficient, reliable and as demonstrated, provides excellent accuracy results for low as well as higher 
excitations. Both $s$-wave and rotational states are considered for four diatomic molecules. All the nine states belonging to vibrational
quantum number $n \leq 2$ and rotational quantum number $\ell \leq 2$ are calculated. Also many higher lying states with $\ell=10,20,25$ have also 
been studied. Due comparisons are made wherever possible. Whenever available, our eigenvalues are either practically identical or competitive to the 
most accurate values \cite{nasser07} in the literature as of now. Otherwise, energies in all cases are found to be noticeably 
superior to all other hitherto reported results. A thorough analysis is presented on changes in energy with respect to state indices $n,\ell$. 
Several new states are reported. In view of the simplicity and accuracy offered by this approach for one of the most fundamental model molecular 
potentials, it is hoped that the scheme will be equally successful and useful for other relevant potentials in molecular physics and related areas.

\section{acknowledgment} It is a pleasure to thank the IISER-Kolkata colleagues for many fruitful discussions. I also thank the referee for his
kind comments and suggestions.


\begin{thebibliography}{99}
\bibitem{morse29} P.~M.~Morse, Phys.~Rev.~ \textbf{34}, 57 (1929). 
\bibitem{popov01} D.~Popov, Phys.~Scripta \textbf{63}, 257 (2001). 
\bibitem{dong02} S.-H.~Dong, R.~Lemus and A.~Frank, Int.~J.~Quant.~Chem.~ \textbf{86}, 433 (2002). 
\bibitem{dong02a} S.-H.~Dong, Can.~J.~Phys.~ \textbf{80}, 129 (2002). 
\bibitem{dong06} D.~Popov, I.~Zaharie and S.-H.~Dong, Czechoslovak J.~Phys.~ \textbf{56}, 157 (2006). 
\bibitem{dong03} S.-H.~Dong and G.-H.~Sun, Phys.~Lett.~A \textbf{314}, 261 (2003). 
\bibitem{dong03a} S.-H.~Dong, Y.~Tang and G.-H.~Sun, Phys.~Lett.~A \textbf{320}, 145 (2003). 
\bibitem{yu04} J.~Yu, S.-H.~Dong and G.-H.~Sun, Phys.~Lett.~A \textbf{322}, 290 (2004). 
\bibitem{serrano10} F.~A.~Serrano, X.-Y.~Gu and S.-H.~Dong, J.~Math.~Phys.~ \textbf{51}, 082103 (2010).
\bibitem{dong12} G.-H.~Sun and S.-H.~Dong, Comm.~Theor.~Phys.~ \textbf{58}, 815 (2012).
\bibitem{duff78} M.~Duff and H.~Rabitz, Chem.~Phys.~Lett.~ \textbf{53}, 1 (1978). 
\bibitem{depristo81} A.~E.~DePristo, J.~Chem.~Phys.~ \textbf{74}, 9 (1981). 
\bibitem{elsum82} I.~R.~Elsum and R.~G.~Gordon, J.~Chem.~Phys.~ \textbf{76}, 5452 (1982).  
\bibitem{morales89} D.~A.~Morales, Chem.~Phys.~Lett.~ \textbf{161}, 253 (1989).
\bibitem{bag92} M.~Bag, M.~M.~Panja, R.~Dutt and Y.~P.~Varshni, Phys.~Rev.~A \textbf{46}, 6059 (1992).
\bibitem{filho00} E.~D.~Filho and R.~M.~Ricotta, Phys.~Lett.~A \textbf{269}, 269 (2000). 
\bibitem{morales04} D.~A.~Morales, Chem.~Phys.~Lett.~ \textbf{394}, 68 (2004). 
\bibitem{pekeris34} C.~L.~Pekeris, Phys.~Rev.~ \textbf{45}, 98 (1934). 
\bibitem{berkdemir05} C.~Berkdemir and J.~Han, Chem.~Phys.~Lett.~ \textbf{409}, 203 (2005). 
\bibitem{castro06} E.~Castro, J.~L.~Paz and P.~Mart\'in, J.~Mol.~Struct.~THEOCHEM \textbf{769}, 15 (2006).
\bibitem{bayrak06} O.~Bayrak and I.~Boztosun, J.~Phys.~B \textbf{39}, 6955 (2006). 
\bibitem{aldossary07} O.~M.~Al-dossary, Int.~J.~Quant.~Chem.~ \textbf{107}, 2040 (2007).
\bibitem{nasser07} I.~Nasser, M.~S.~Abdelmonem, H.~Bahlouli and A.~D.~Alhaidari, J.~Phys.~B \textbf{40}, 4245 (2007). 
\bibitem{qiang07} W.~-C.~Qiang and S.~-H.~Dong, Phys.~Lett.~A \textbf{363}, 169 (2007). 
\bibitem{ikhdair09} S.~M.~Ikhdair, Chem.~Phys.~Lett.~ \textbf{361}, 9 (2009). 
\bibitem{leykoo95} E.~Ley-Koo, S.~Mateos-Cort\'es and G.~Villa-Torres, Int.~J.~Quant.~chem. \textbf{56}, 175 (1995). 
\bibitem{taseli98} H.~Ta\c seli, J.~Phys.~A \textbf{31}, 779 (1998). 
\bibitem{barakat06} T.~Barakat, K.~Abodayeh, O.~M.~Al-Dossary, Czechoslovak J.~Phys.~ \textbf{56}, 583 (2006). 
\bibitem{roy04} A.~K.~Roy, Phys.~Lett.~A \textbf{321}, 231 (2004).
\bibitem{roy04b} A.~K.~Roy, J.~Phys.~B \textbf{37}, 4369 (2004); {\it ibid.} \textbf{38}, 1591 (2005).
\bibitem{roy05} A.~K.~Roy, Int.~J.~Quant.~Chem.~\textbf{104}, 861 (2005).
\bibitem{roy05a} A.~K.~Roy, Pramana--J.~Phys.~ \textbf{65}, 01 (2005). 
\bibitem{roy07} A.~K.~Roy and A.~F.~Jalbout, Chem.~Phys.~Lett.~ \textbf{445}, 355 (2007). 
\bibitem{roy08} A.~K.~Roy, A.~F.~Jalbout and E.~I.~Proynov, Int.~J.~Quant.~Chem.~ \textbf{108}, 
827 (2008). 
\bibitem{roy08a} A.~K.~Roy, A.~F.~Jalbout and E.~I.~Proynov, J.~Math.~Chem.~ \textbf{44}, 260 (2008). 
\bibitem{roy11} A.~K.~Roy, in \emph{Mathematical Chemistry}, W.~I~Hong (Ed.), Nova Science Publishers, 
Hauppauge, NY, USA, pp.~555-599 (2011). 
\bibitem{roy13} A.~K.~Roy, Int.~J.~Quant.~Chem.~\textbf{113}, 1503 (2013).
\bibitem{nist} The NIST reference on constants, units and uncertainty, physics.nist.gov/cuu/Constants/index.html.
\end{thebibliography}
\end{document}